\documentclass{kapproc} % Computer Modern font calls
\begin{document}
\input epsfig.sty
%%   of the \usepackage commands below:
%%%%%
%% If you use a font encoding package, please enter it here, i.e.,
%% \usepackage{T1enc}
%%%%%
%  If you have MathTimes and MathTimesPlus fonts, you
%  may uncomment the line below and use them, but you are
%  not obligated to do so, and most authors do not have
%  these fonts. (You may need to edit m-times.sty to make the
%  font names match those on your system)
%  You must have the MathTimes fonts for this to work. They may be
%  purchased from the Y&Y company, http://www.YandY.com.
%% \usepackage[mtbold,noTS1]{../mymtimes}
%%%%%
% PostScript font calls
%
% If you use the procps PS font file, you may need to edit it
% to make sure the font names match those on your system. See
% the top of the procps.sty file for more info.
%\usepackage{procps} 
%%%%%%%%%%%%%%%%%%%%%%%%%%%%%%%%%%%
%% LaTeX209, uncomment only one:
%% (Make sure documentclass and usepackage commands above are commented out!)
%\begin{document}
%\documentstyle{kapproc} % Computer Modern fonts
%  \documentstyle[procps]{kapproc} %For PostScript fonts
%   (MathTimes style is not available for authors using LaTeX2.09)
%%%%%%% Formatting Commands You Can Set or Change ===>>>
%  optional, uncomment to make current time and `draft' appear at
%  bottom of page.
%\draft
%%%% To change footnotes to appear at bottom of page ==>
%% (Default is endnotes that appear at the end of the chapter, above
%%   the references, or whereever \notes is written.)
%% uncomment to make footnote appear at bottom of page:
\let\footnote\savefootnote
%% uncomment if you want footnotetext to appear at the bottom of the page:
\let\footnotetext\savefootnotetext 
%% uncomment if you want a ruled line above the footnote:
\let\footnoterule\savefootnoterule 
%%%% <== end footnote changes
%% How many levels of section head would you like numbered?
%% 0= no section numbers, 1= section, 2= subsection, 3= subsubsection
%%==>>
%\setcounter{secnumdepth}{3}
%% How many levels of section head would you like to appear in the
%% Table of Contents?
%% 0= chapter titles, 1= section titles, 2= subsection titles,
%% 3= subsubsection titles.
%%==>>
\setcounter{tocdepth}{1}
%%%%%%% Bibliography Style Settings ==>>
%%% Uncomment one of the Following:
\kluwerbib
%\normallatexbib
%%%%%%%
% \kluwerbib will produce this kind of bibliography entry:
%
% Anderson, Terry L.,...
%   More bib entry here...
%
% \cite{xxx} will print without brackets around the citation.
%
 \bibliographystyle{apalike} %should be use with \kluwerbib
%%%%%%%
% \normallatexbib will produce bibliography entries as shown in the
% LaTeX book
%
% [1] Anderson, Terry L.,...
%     More bib entry here...
%
%\cite{xxx} will print with square brackets around the citation, [1].
%
% Any \bibliographystyle{} may be used with \normallatexbib, but
% you should check with your editor to find the style preferred for
% the book you are contributing to.
%%%%%%% To change brackets around citation ==>>
% Default with \kluwerbib is no brackets around citation. 
% Default with \normallatexbib is square brackets around citation.
%If you want parens, around citation, i.e., (citation), uncomment these lines:
%\let\lcitebracket(
%\let\rcitebracket)
%%%%%%%  <<== End Bibliography Style Settings
%%%%%%% Author and Topic Indices
%% If you want to have both an author and a topic index, uncomment this:
%\startauthorindex
%%%% <<== End Formatting Commands You Can Set or Change %%%%%%%%%%%%%%%%%
%%%%%%%%%%%%%%%%%%%%%%%%%%%%%%%%%%%%%%%%%%%%%%%%%%%%%%%%%%%%%%%%%%%%%%%%%
%\begin{document}
%------------ article title  ------------------->>
% If you use \\'s , please supply an alternate version of the title
% in square brackets, i.e., 
\articletitle{Polymer chain collapse in supercritical fluids}
%\articletitle[]{}
%% optional, to supply a shorter version of the title for the running head:
%%\chaptitlerunninghead{}
\author{C. Ortiz-Estrada$^{a,b}$, 
G. Luna-B\'arcenas$^{c,}$\footnote{gluna@arcos.qro.cinvestav.mx},
G. Ram\'\i rez-Santiago$^{d,e}$, 
I. C. Sanchez$^{e,}$\footnote{ sanchez@che.utexas.edu}
and J. F. J. Alvarado$^{a}$\\
$^{a}$ Chemical Engineering Department, Instituto Tecnologico
de Celaya, Celaya Guanajuato, 38010 MEXICO\\
$^{b}$ ESQIE, Instituto Polit\'ecnico Nacional and 
Universidad Iberoamericana A. C., 
M\'exico D. F. MEXICO\\
$^{c}$ Laboratorio de Investigaci\'on en Materiales, CINVESTAV
Unidad Que\-r\'e\-taro, Que\-r\'e\-taro 76230 MEXICO\\
$^{d}$ Instituto de F\'\i sica, UNAM, PO Box 20-364, M\'exico 
01000 D. F.
, MEXICO\\
$^{e}$ Chemical Engineering Department, The University of Texas at 
Austin, Austin TX 78712 U.S.A. }
\begin{abstract}
Recent  computer simulations of a polymer chain in a solvent have 
provided evidence, 
for the first time, of polymer chain collapse near 
the lower critical solution temperature (LCST). Motivated by these results, 
we have studied further this system
to understand the effect of solvent and monomer sizes, chain length, and solvent and 
monomer energetic interactions. By means of extensive Monte Carlo simulations, the mean radius 
of gyration $R_{g}$ and end-to-end distance  $R$, 
are calculated for a single chain in a solvent over a broad range 
of volume fractions, pressures and temperatures. Our results indicate that
in general, the chain collapses as 
temperature increases at constant pressure, or as density decreases at constant temperature. 
A minimum in $R_{g}$ and $R$ occurs near the LCST and slightly 
above the coil-to-globule transition 
temperature (C-GTT), where the chain adopts a quasi-ideal conformation, defined by the balance 
of binary attractive and repulsive interactions. At temperatures 
well above the LCST, the chain 
expands again suggesting an upper critical solution temperature (UCST) 
phase boundary above the 
LCST forming a closed-immiscibility loop. However, this observation 
strongly depends on the 
solvent-to-monomer size ratio.
\end{abstract}
% optional keywords
%------------ body of article ------------------->>
\section{Introduction}
As it is well known, polymer solutions also exhibit lower critical 
solution temperature
(LCST) or thermally induced phase separation when temperature is raised. 
LCSTs have been
observed in strongly interacting polar mixtures, for example aqueous solutions, as well 
as in weakly interacting non-polar polymer solutions. An LCST may also be produced by the
mixture's finite {\it compressibility}. Unlike the upper critical solution temperature
(UCST), which is driven by unfavorable energetics, a thermodynamics 
analysis shows that
LCST is an entropically driven phase separation.

LCSTs usually occur in the vicinity of the vapor-liquid critical 
temperature of the pure
solvent. Specifically, they are typically observed at about 
0.7$T^{*}_{c}$ to 0.9$T^{*}_{c}$.
Of particular relevance is the phase behavior 
of polymer solutions
in supercritical fluids (SCFs), since most of these systems exhibit an LCST. 

The development of a fundamental understanding of the phase behavior of polymer solutions
in SCFs is a theoretical challenge of great practical interest[1]. 
The solution
behavior is complex due to large values of free volume, isothermal compressibility, volume
expansivity and concentration fluctuations. Practical SCF applications involving LCST phase
behavior include polymer fractionation, impregnation and purification, polymer extrusion
and foaming, formation of materials by rapid expansion from supercritical solution and
precipitation with a compressed fluid antisolvent, dispersion as well as
emulsion polymerization,
and formation of emulsions and microemulsions [2,3].

It was conjectured some time ago that chain collapse should also 
be observed near an LCST
in an analogous way as it occurs near an UCST [4]. The implications of this
polymer physics problem are of great relevance to SCF technology. 
In 1997, Luna-B\'arcenas,
{\it et. al.} [5], reported for the first time, from extensive 
numerical calculations, evidence
of polymer chain collapse near the LCST upon heating the polymer 
solution at constant
pressure. By investigating single chain architecture, the phase behavior 
of more concentrated
polymer solutions was predicted. In other words, the physics of chain 
collapse near LCST
captures the macroscopic phase separation behavior in a finite concentration polymer
solution. Luna-B\'arcenas, {\it et. al.} observed that the collapse of a single
polymer chain correlates well with {\it coil-globule transition 
temperature (CGT-T)} and an
occurrence of an LCST phase boundary. It was also observed that upon further heating the
collapsed chain expanded again suggesting the presence of a closed immiscibility loop.
Later, this closed loop was corroborated by direct phase separation simulations using
an expanded Gibbs ensemble formalism [5]. However, the above studies were
restricted to a fixed chain length, energetic interactions and solvent to 
monomer size ratio. 
Motivated by these findings, in this work we report some results of more 
extensive numerical simulations by investigating
the effect, that  energetic interactions and monomer to solvent size,
have  on the phase diagram of a mixture.

\section{Molecular model}
The system studied in this work consists of a single freely jointed chain 
inmersed in a solvent medium and is analogous to the infinite
dilute regime of a polymer system. 
This regime consists of a polymer system in which
chains are far from each other to avoid interchain  interactions, 
that is, the chains act as individual entities.
To simulate the intermolecular interactions  we used 
the typical Lennard-Jones (LJ) potential that is defined as,
\begin{eqnarray}
U_{i,j}(r) &=& 4\epsilon_{ij}\Big[ \big({\sigma_{ij}\over {r}}\big)^{12}
 - \big( {\sigma_{ij}\over {r} }\big)^{6} \Big] + 0.01632\epsilon_{ij},
\hskip 0.75in {\rm if} \>\>\> r \leq  2.5\sigma_{ij}.\cr
              &  & 0 \hskip 3.24in {\rm otherwise.} \nonumber
\end{eqnarray}
In this equation $r$ is the distance between molecules, 
$\sigma_{ij}$ and $\epsilon_{ij}$
represent the parameters of the potential. Since we have introduced
a cutoff of $2.5\sigma$ in the interaction LJ potential, 
it is equal to zero for larger distances. This is equivalent to 
an upwards shift in the entire potential . The phase and critical
behavior of this LJ model have been studied by 
Smit [6] who reported
a reduced critical temperature $T^{*} =  ( K_{B}T_{c} / \epsilon )=1.08$, 
a reduced
critical density $\rho^{*} = \sigma^{3}\rho_{c} = 0.31$, and a reduced 
critical pressure
$P^{*} = P_{c}\sigma^{3} / \epsilon = 0.10$. In this paper, we 
report results corresponding to a 
constant chain length of $N=20$ segments or monomers. The strength of 
the energetic
interactions has been varied by changing sistematically  the ratio 
$\epsilon_{11}/\epsilon_{22}$.
Here, the subscript 11 refers to the nonbonded monomer-monomer 
interaction while the
subscript 22 is related to the solvent-solvent interaction.
In the same manner, the monomer-solvent size effect is considered 
by varying the ratio 
$\sigma_{11}/\sigma_{22}$, with $\sigma_{11}$ the monomer size 
and $\sigma_{22}$ the
solvent size. The way the site density is usually defined, 
($\rho^{*}=\sigma^{3}\rho$)
is not appropiate when dealing with objects of dissimilar sizes. 
Instead, we use the
volume fraction, $\eta=(site\>\> volume)/(total\>\> volume)$ as a more natural 
variable since it
takes into account the volume of the objects, whereas the site 
density does not. 

To study the polymer chain collapse we used the continuum configurational bias 
(CCB) Monte Carlo algorithm. This method consists in cutting the chain
at a random site. A portion of the chain is then deleted from this 
site to one of the ends of the chain.
Finally, the chain  is regrown site by site until its original 
length is restored. A more detailed description and explanation
of this algorithm is presented in reference [7].

\section{Results and discussion}
\subsection{Solvent-to monomer size ratio effect:(small solvent-big monomer)}
To understand better the effect of the monomer-to-solvent size ratio on the 
mixture's phase behavior, we considered a monomer segment volume that
is twice the solvent volume. This is equivalent to consider  a ratio
$\sigma_{11}/\sigma_{22} = 1.26$, 
The energetic interactions
were chosen such that the interactions between monomer-monomer 
and solvent-solvent are equal,
that is, $\epsilon_{11}/\epsilon_{22}=1$ keeping this ratio constant.
\begin{figure}[t]
\epsfxsize=9cm
\epsfysize=8cm
\centerline{\epsffile{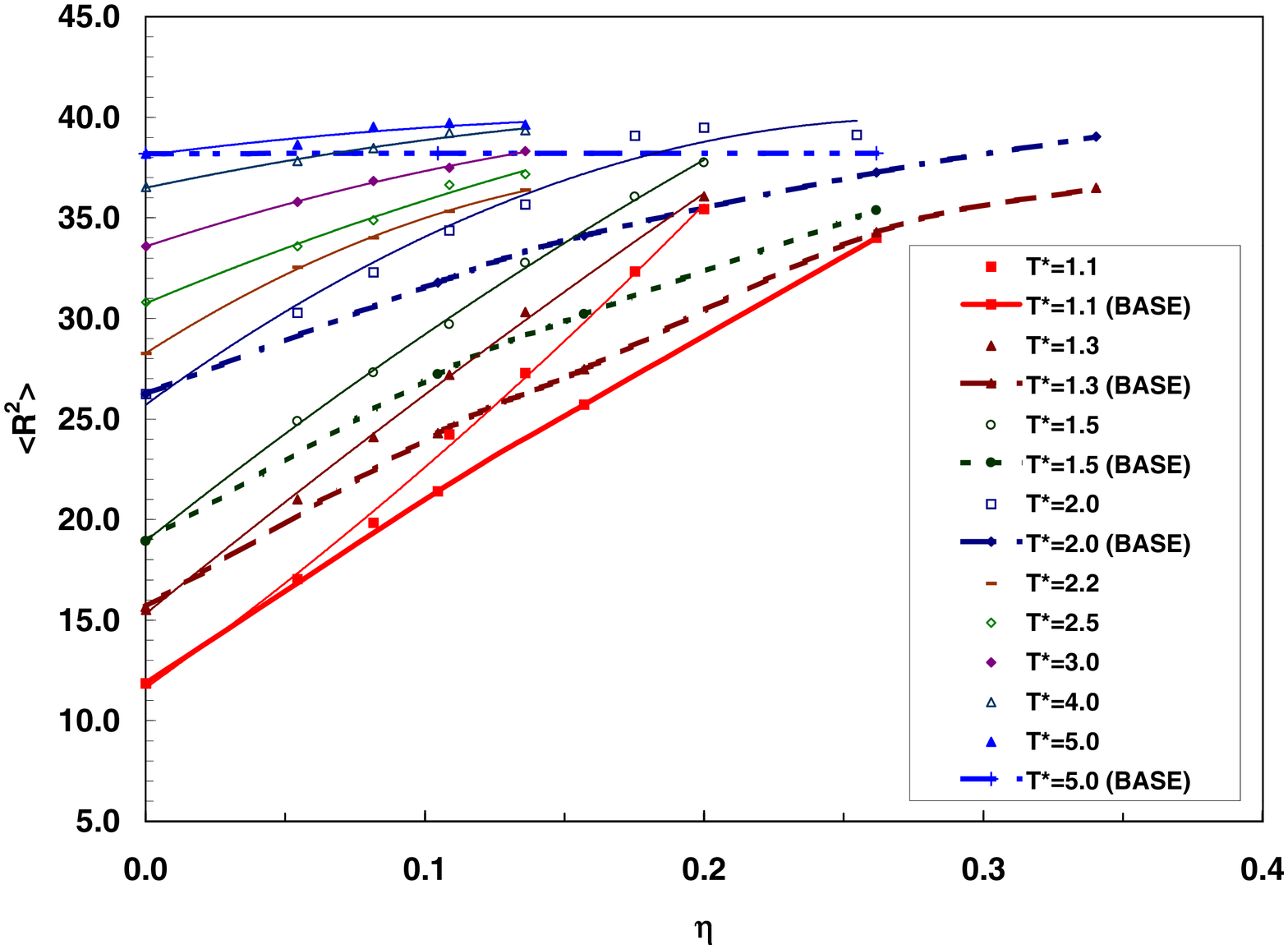}}
\makebox[1cm]{}
\caption{Mean square end-to-end distance versus volume fraction for
a small solvent-big polymer system. See text for a discussion
of this figure.}
%\label{fig-1}
\end{figure}
\begin{figure}[b]
\epsfxsize=9cm
\epsfysize=8cm
\centerline{\epsffile{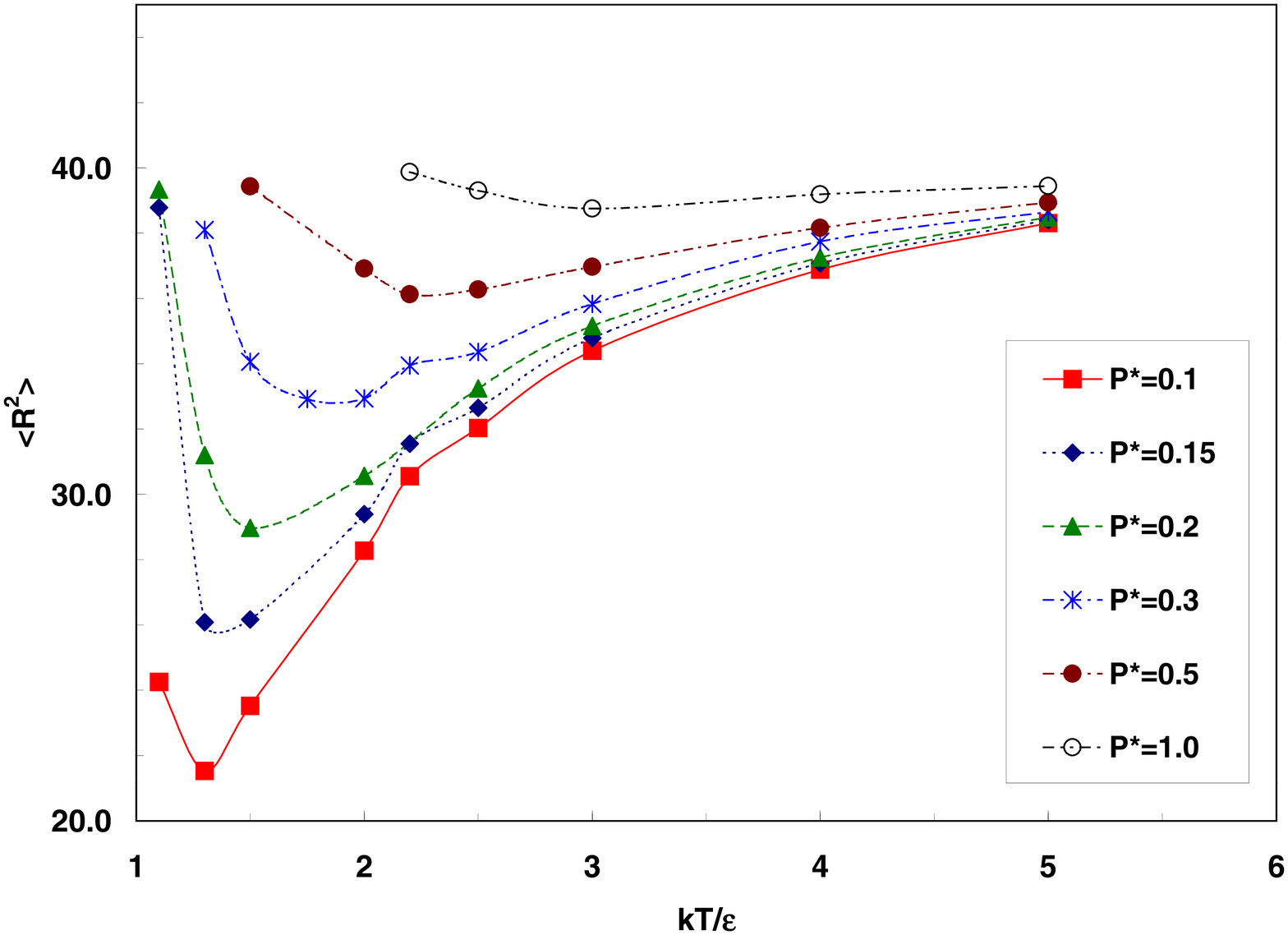}}
\makebox[1cm]{}
\caption{Mean square end-to-end distance versus temperature for
a small solvent-big polymer system. These results are discussed in
the text.}
%\label{fig-2}
\end{figure}
Figure 1 shows the mean square end-to end 
distance $<R^{2}>$ of the chain as a function of
the system volume fraction at several reduced temperatures, that were 
chosen in the vicinity of the
pure solvent critical temperature $T^{*}_{c}=1.08$. For reasons of 
comparison it is important to note that in figure 1
the data labeled with the legend "BASE" represent the results 
for a symmetric
mixture, that is, $\epsilon_{11}/\epsilon_{22}=1$ and 
$\sigma_{11}/\sigma_{22} = 1$, as
reported in reference [5]. At high densities, the chain adopts a 
coil-like conformation that approaches the athermal or infinite 
temperature limit.
The chain collapse at low densities suggest that solvent quality 
diminishes as the solvent
density or volume fraction decreases. This behavior is experimentally 
observed in a pure
supercritical fluid. For instance, the square of the solubility 
parameter (cohesive energy
density per unit volume) decreases when the density also decreases.
It is interesting to note that going from low to high volume 
fraction the {\it small 
solvent-big polymer} mixture expands more rapidly than the symetric 
mixture at  fixed
temperature. This suggest that a smaller solvent with similar 
energetic interactions 
compared to a bigger one acts as a {\it better solvent} since the chain is more solvated.
Also note from Figure 1 that the increase in chain dimensions 
with temperature at constant
volume fraction --pressure must also increase to mantain 
constant the density-- is consistent 
with the idea that attractive energetics become less important at 
high temperatures, that
is, $\epsilon/(k_{B}T) \rightarrow 0$. When the chain collapses to 
enhance favorable
intrachain attractive interactions, it does so at the expense of losing 
chain conformational
entropy. Chain connectivity brings chain segments into close proximmity 
to one another
--the so called correlation hole-- enhancing the effects of intrachain 
forces relative to
chain-solvent interactions. The remarkable 
behavior of chain collapse
with temperature is shown in figure 2. 
At constant pressure chain dimensions go through a
minimum suggesting that a phase boundary is being approached. It has been shown 
recently [5] that this minimum represents the LCST of the mixture.
However, in our case {\it small solvent-big polymer}, the chain's 
minimum dimension
is bigger than the symmetric case, that is, {\it same size solvent and monomer}, which is
in agreement with the observation made above from figure 1, that
smaller solvent solvates better
the chain. The expansion of the chain upon further heating ---see figure 2---
suggest the existence
of a one-phase region. These observations were already pointed out in
reference [5]. In fact they demonstrated, by direct simulation of phase 
equilibria,
the existence of a closed immiscibility loop in the polymer solvent 
phase diagram.
\begin{figure}[t]
\epsfxsize=9cm
\epsfysize=8cm
\centerline{\epsffile{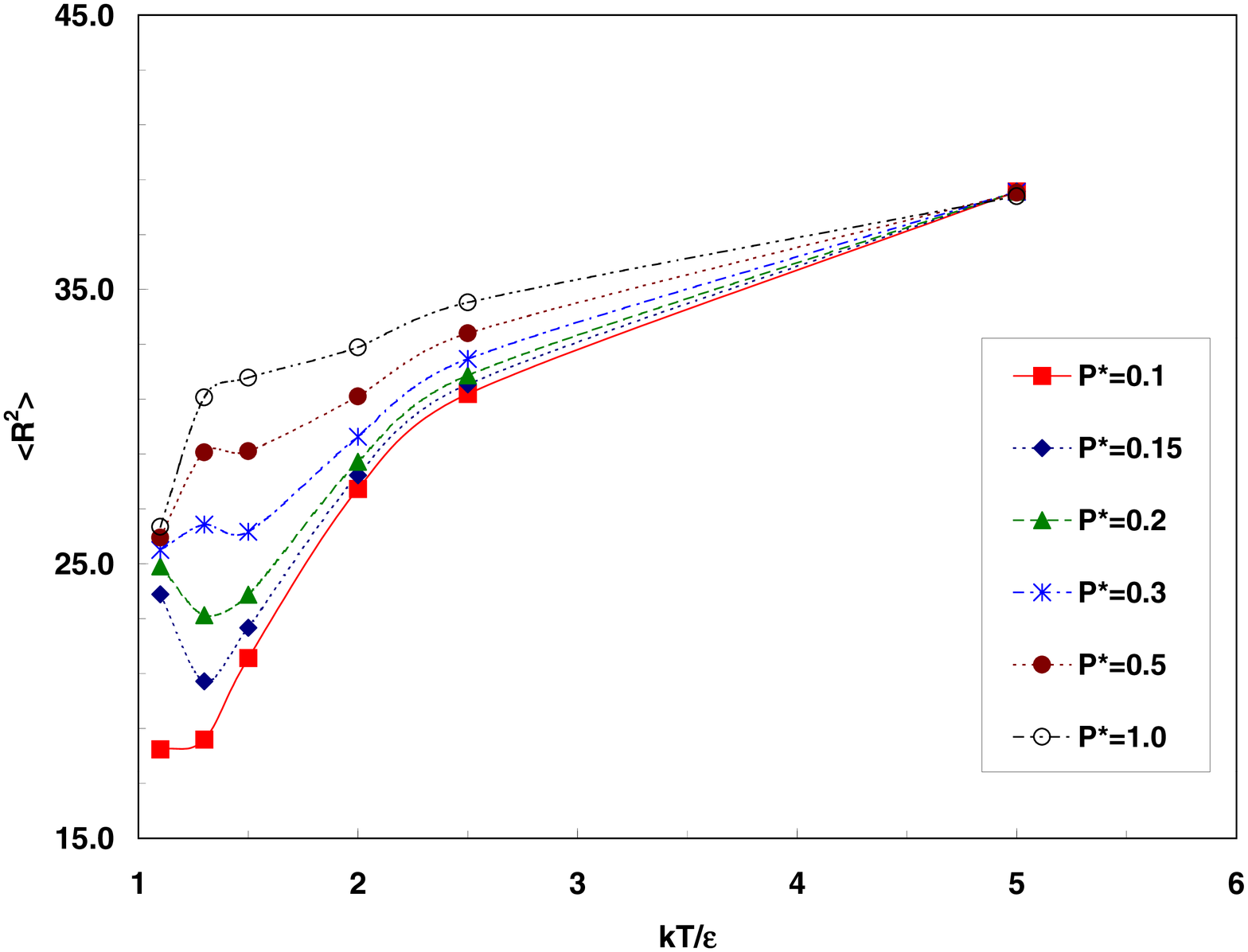}}
\makebox[1cm]{}
\caption{Mean square end-to-end distance versus temperature for
a big solvent-small polymer system. See text for an interpretation
of this figure.}
%\label{fig-3}
\end{figure}
\subsection{Solvent-to-monomer size ratio effect (big solvent small 
polymer case)}
In contrast to the previous subsection, we now consider the effect of a 
solvent that has
twice the volume of the monomer unit in the chain, that is 
$\sigma_{11}/\sigma_{22}
=0.794$. In figure 3 we show the behavior of the mean square end-to-end distance
as a function of temperature. It is worth noting that at a given temperature and
pressure, but at temperatures near the
solvent's critical point, the chain collapses when compared to the {\it small 
solvent-big polymer case}. 
This fact would imply that the mixture would be phase
separated until high enough temperatures are reached at which the mixture will
again be miscible. The predicted phase behavior 
is shown in figure 4 on a P-T plane.
Figure 4 also shows the effect of different energetic interactions. 
\begin{figure}[b]
\epsfxsize=9cm
\epsfysize=8cm
\centerline{\epsffile{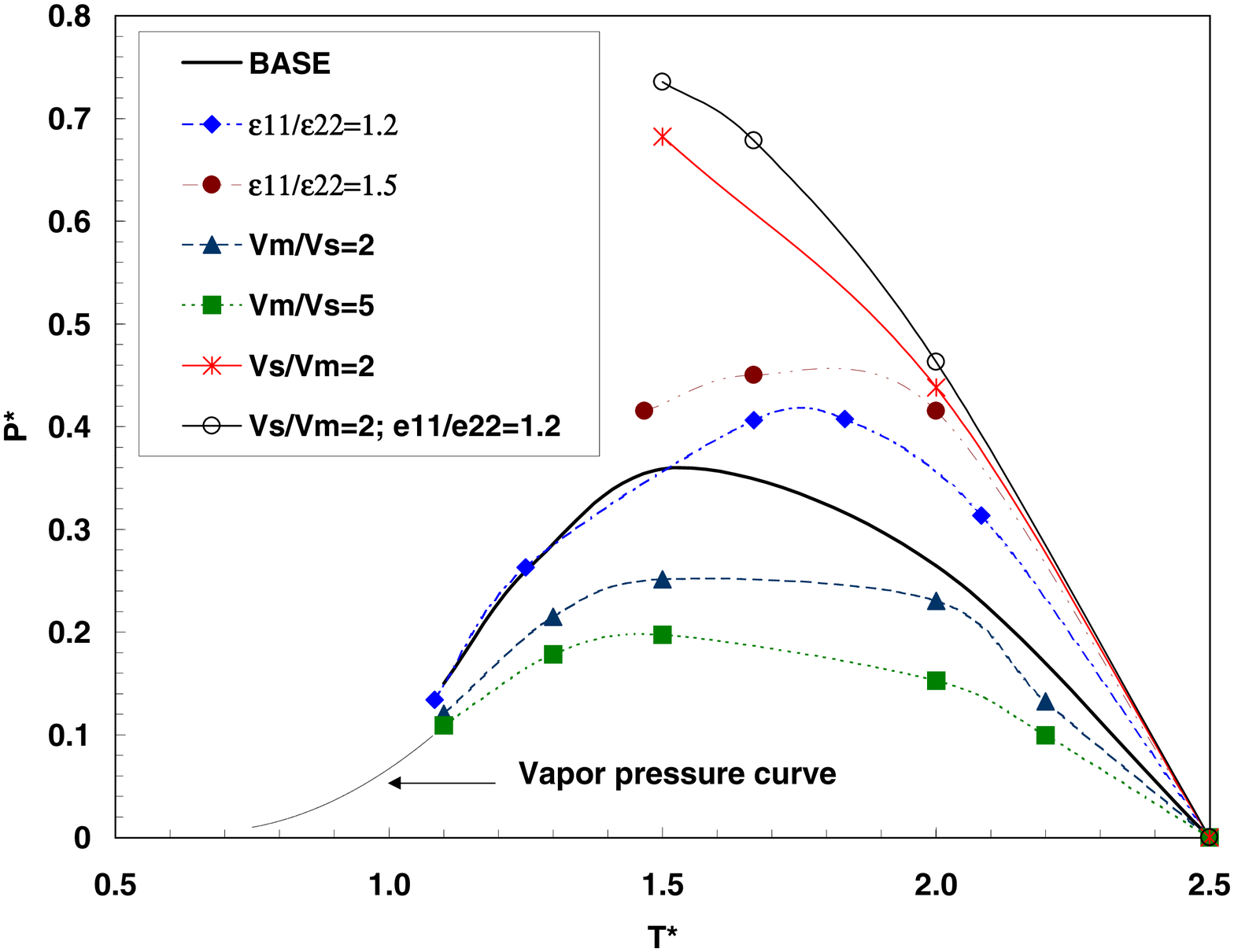}}
\makebox[1cm]{}
\caption{Predicted pressure versus temperature phase diagram.}
%\label{fig-4}
\end{figure}
\begin{figure}[t]
\epsfxsize=9cm
\epsfysize=8cm
\centerline{\epsffile{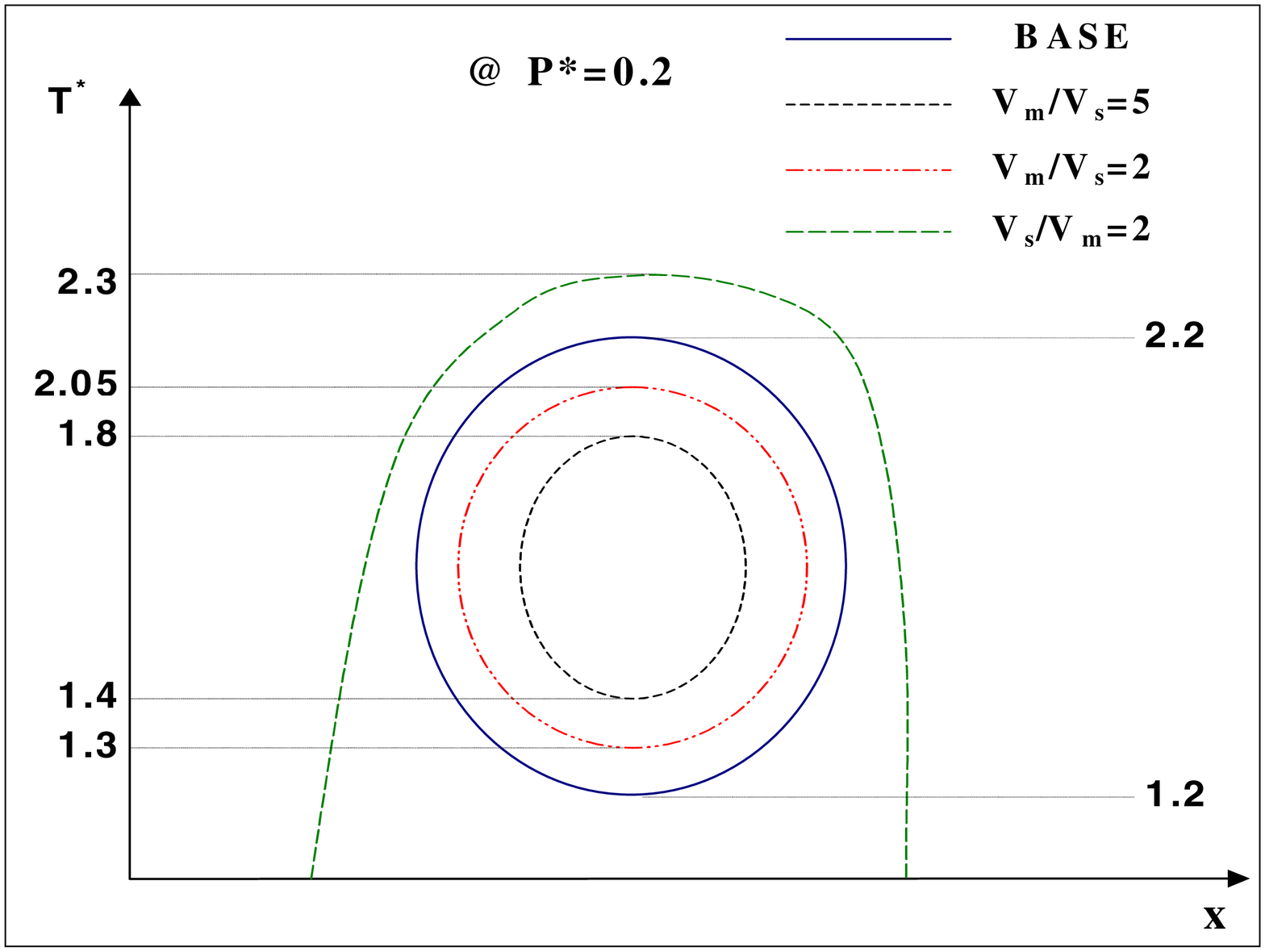}}
\makebox[1cm]{}
\caption{Expected temperature versus composition diagram. The
reduced pressure of the system is $P^{*}=0.2$. }
%\label{fig-5}
\end{figure}
Two cases are
considered: $\epsilon_{11}/\epsilon_{22}=1.2$ and  
$\epsilon_{11}/\epsilon_{22}=1.5$.
In both cases, the non-bonded monomer-monomer energetic interactions 
are stronger
than the solvent-solvent and monomer-solvent interactions. These conditions 
mimic a {\it worse} solvent case with respect to the symmetric mixture.

Finally, a few words on the polymer solution phase diagram. The collapsing chain
--emulating the infinite dilution-- signals the ocurrence of an 
LCST phase boundary in a more
concentrated (finite) solution. Further heating favors the 
expanding of the chain,
which also indicates the existence of a nonhomogeneous region. The macroscopic
picture of the above structural changes --collapsing-expanding-- of the chain  
results in a closed immiscibility loop of varying size and shape. For instance,
the {\it small solvent-big polymer} case predicts a smaller inmiscibility region
whereas the {\it big solvent-small polymer} case predicts a bigger immiscibility
window. For illustrative purposes, the {\it expected phase diagram} for the
different cases explained  above is depicted in figure 5.
%------------ end of article ------------------->>
%% optional
%\section{Summary}
%% optional
\begin{acknowledgments}
Financial support of this work has been  provided by CONACyT, Grants
J33262 (G. Luna-B\'arcenas) and 25298-E and G32723-E 
(G. Ra\-m\'\i rez-Santiago).
We also received financial support from  JIRA-CIN\-VES\-TAV and the NSF Science
\& Technology Center for the Utilization of Carbon Dioxide 
in Manufacturing at the
University of Texas at Austin.
\end{acknowledgments}

%% appendix optional
%\appendix{This is the Appendix Title}
%This is an appendix with a title.
%\appendix{}
%This is an appendix without a title.
% Bibliography made with BibTeX:
%% apalike is preferred if you have used \kluwerbib, above.
%% Otherwise you may use any .bst style your editor approves.
%This will allow many Bib\TeX\ bibliographies in one book.
%See the documentation, edbk.doc, for more information.
\bibliographystyle{apalike}
%\chapbblname{<name of .bbl file>}
%\chapbibliography{<name of .bib file>}
%or 

-----
\end{document}